\documentclass[%
 preprint,
nofootinbib,
 amsmath,amssymb,
 aps,
 prl,
 longbibliography,
 lengthcheck,%
]{revtex4-1}

\usepackage{graphicx}
\usepackage{dcolumn}
\usepackage{bm}
\usepackage{hyperref}
\usepackage{mathtools}
\usepackage[mathlines]{lineno}

\newcommand{\be}{\begin{equation}}
\newcommand{\ee}{\end{equation}}

\newcommand{\la}{\langle}
\newcommand{\ra}{\rangle}

\newcommand{\ssigma}{\bm{\sigma}}

\newcommand{\cH}{{\cal H}}
\newcommand{\cS}{{\cal S}}
\newcommand{\cB}{{\cal B}}

\newcommand{\psiS}{\psi^{\cal S}}

\newcommand{\densS}{ \rho^{\cal S}}

\newcommand{\meandensS}{\overline{{ \rho}^{\cal S}}}

\newcommand{\tr}{\mathrm{tr}}

\newcommand{\ii}{\mathrm{i}}

\begin{document}


\title{
Decoherence at the level of eigenstates
}

\author{Oleg Lychkovskiy}

\affiliation{Skolkovo Institute of Science and Technology,
Skolkovo Innovation Center 3, Moscow  143026, Russia}
\affiliation{Steklov Mathematical Institute of Russian Academy of Sciences,
Gubkina str. 8, Moscow 119991, Russia}

\date{\today}


\begin{abstract}
An eigenstate decoherence hypothesis states that each individual eigenstate of a large closed system is locally classical-like. We extend this hypothesis to account for a typically extremely short time scale of decoherence. The extension implies that nondiagonal matrix elements of certain operators -- quantumness witnesses -- are suppressed as long as the energy difference between corresponding eigenstates is smaller than the inverse decoherence time.
\end{abstract}

\maketitle

\noindent{\it Introduction.}
Superposition principle of quantum theory implies the existence of a vast body of states which have never been observed - such as a state of a cat being alive and dead simultaneously,  or a state of a soccer ball being situated in different corners of the pitch simultaneously.  The decoherence programm pioneered by Zeh \cite{zeh1970interpretation,zeh1973toward} and Zurek \cite{zurek1981pointer,zurek1982environment} suggests a solution of this conundrum: One notes that inevitable interaction of macroscopic bodies  with their environments makes ``nonclassical'' superposition states very fragile. Even if such a nonclassical state were created, on a very short time scale $\tau$ it would dynamically ``decohere'', i.e. entangle with the environment in a way that effectively erases all nonclassical features \cite{schlosshauer2008decoherence}. The same interaction will lead to  thermalization on a longer time scale $T$ \cite{gemmer2009quantum}. For macroscopic and even mesoscopic systems $\tau$ is smaller than $T$ by many orders of magnitude.
Thanks to this drastic  separation of time scales, one can distinguish decoherence from thermal equilibration and regard them as separate phenomena.

An important role in understanding thermalization is played by the eigenstate thermalization hypothesis (ETH) which essentially asserts that an individual eigenstate of a large closed quantum system is locally thermal. \cite{deutsch1991quantum,srednicki1994chaos,rigol2008thermalization} The present author proposed a completely analogous eigenstate decoherence hypothesis (EDH) some time ago \cite{lychkovskiy2013dependence} which stated that an individual eigenstate is locally classical-like. That version of EDH did not address, however, a crucial feature of decoherence -- its extremely short time scale.  Here we extend the EDH to fill this gap.

\smallskip
\noindent{\it Preliminaries.}
Consider a closed quantum system  $\cH$  which is divided in two parts, a subsystem $\cS$ and the environment (bath), $\cB$,
\be
\cH=\cS\otimes\cB,
\ee
where we use the same notations, $\cH$, $\cS$ and $\cB$, for the corresponding Hilbert spaces. The state of the whole system is described by the time-dependent state vector $\Psi_t\in\cH$. It evolves according to the Schrodinger equation
\be
\ii \partial_t \Psi_t = H \Psi_t,
\ee
where $H$ is the Hamiltonian of the closed system, and we put $\hbar=1$. The subsystem $\cS$ is described by the density matrix
\be\label{densS}
\densS_t \equiv \tr_{\cB} |\Psi_t\ra\la\Psi_t|,
\ee
where $\tr_{\cB}$ is the partial trace over the environment $\cB$. Its time evolution is described by
\be\label{densSt}
\densS_t=\sum_{n,m}  c_n c_m^* e^{-\ii (E_n-E_m)t} \densS_{nm},
\ee
where $n$ and $m$ enumerate the eigenstates $\Phi_n$ of the total Hamiltonian,
\be
H \Phi_n =E_n \Phi_n,
\ee
$E_n$ are corresponding eigenenergies,
 the matrices $\densS_{nm}$ are defined as
\be
\densS_{nm} \equiv \tr_{\cB}|\Phi_n\ra\la \Phi_m|,
\ee
and a coefficient $c_n$ is equal  to the overlap between the initial state $\Psi_0$ and the $n$'th eigenstate,
\be
c_n \equiv \la \Phi_n | \Psi_0 \ra.
\ee

\smallskip
\noindent{\it Eigenstate thermalization.} We first briefly remind the eigenstate thermalization hypothesis \cite{deutsch1991quantum,srednicki1994chaos,rigol2008thermalization}.  At large times, $t \sim T$, one expects that exponents $e^{-\ii (E_n-E_m)t}$ average out for nondiagonal entries with $|E_n-E_m|> 2\pi/T$. This idea can be brought to the extreme by considering the time-averaged density matrix of the subsystem,
\be\label{meandensS}
\meandensS\equiv \lim_{t \to \infty}\frac{1}{t}\int_{0}^t \densS_{t'} \, dt'.
\ee
Clearly, in the absence of degeneracies one obtains from eq. \eqref{densSt}
\be\label{meandensS diag}
\meandensS=\sum_{n}  |c_n|^2 \densS_{nn}.
\ee
The eigenstate thermalization hypothesis states that $\densS_{nn}$ is insensitive to the microscopic details of the eigenstate $\Phi_n$ and, in the simples case, is a smooth function of the sole characteristics of the eigenstate -- its energy,
\be
\densS_{nn}=\densS_{eq}(E_n),
\ee
up to corrections irrelevant for large system sizes.

This hypothesis, if valid, automatically ensures one of the main aspects of thermalization -- namely, the independence of the equilibrium state from the microscopic details of initial conditions. Further, if one assumes that the interaction between the subsystem and the environment is sufficiently weak, one can prove that $\densS_{eq}(E)$ has a Boltzmann-Gibbs form. In essence, the ETH ensures thermalization of a subsystem at the level of an individual eigenstate of a closed system consisting of the subsystem and its environment. Various case studies confirmed that $\densS_{nn}$ indeed has all attributes of an equilibrium state \cite{rigol2008thermalization,kim2014testing,beugeling2014finite-size,magan2016random}.


\begin{figure*}[t]
\includegraphics[width = \textwidth]{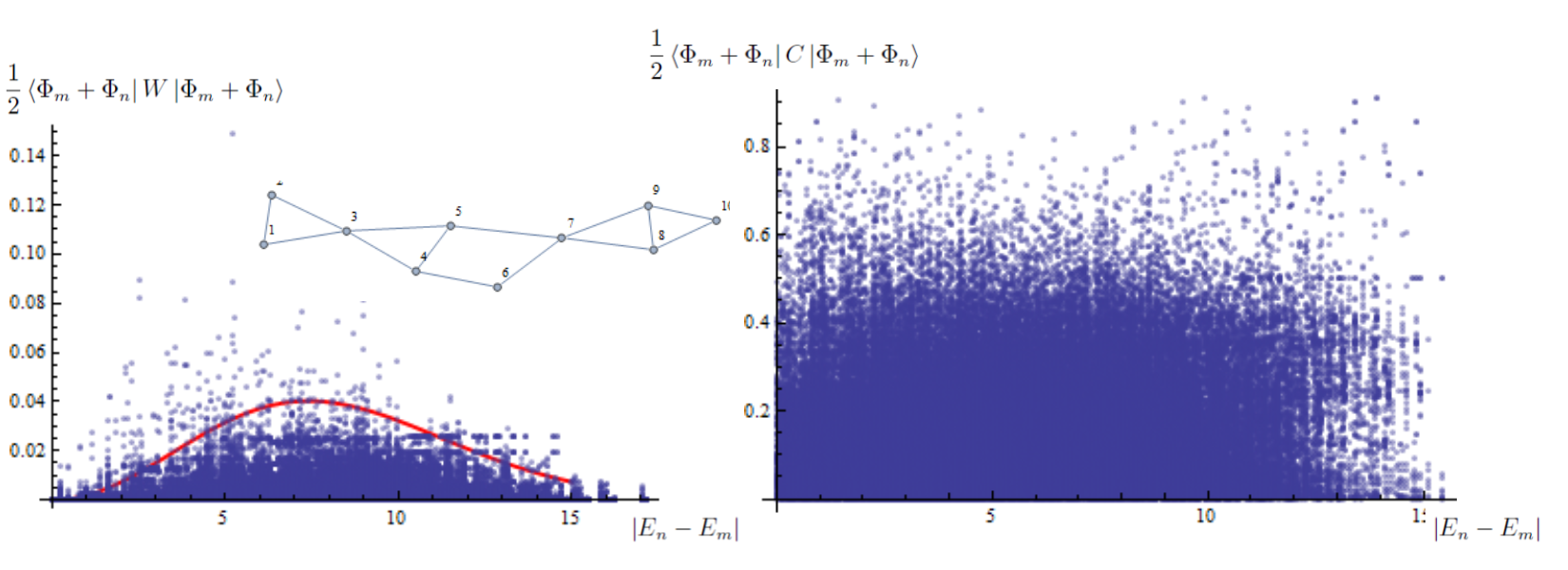}
\caption{Left: quantumness witness \eqref{EDH vulgar} as a function of $|E_n-E_m|$ with an operator $W$ defined in eq. \eqref{W}, for a spin system \eqref{H} on a graph shown in the inset. Red line is a guide for an eye.  Right: analogous matrix elements for an operator $C$ defined in eq. \eqref{C}.
\label{fig}}
\end{figure*}

\smallskip
\noindent{\it Eigenstate decoherence.} As was noted in ref. \cite{lychkovskiy2013dependence}, one reasonably expects from the state $\densS_{nn}$ that it is not only thermal but also classical. This natural idea was referred to as the "eigenstate decoherence hypothesis" (EDH) in  ref. \cite{lychkovskiy2013dependence}, in analogy to the ETH. Here we propose an extension of this hypothesis which accounts for the extremely short decoherence time scale, $\tau$.

However, first we have to introduce some quantitative measure of quantumness of a density matrix. A question how to quantify quantumness attracted a lot of attention and was addressed in various ways, see e.g. Ref. \cite{nimmrichter2013macroscopicity} and references there in. For the purposes of the present paper we employ a technically simple sufficient criterion of quantumness which is constructed in the spirit of the entanglement witness \cite{terhal2000bell,horodecki2009quantum}. Namely, we choose a suitable Hermitian operator $W$ acting in $\cS$ with a unit operator norm and calculate a ``quantumness witness''
$
w(\densS) \equiv |\tr_\cS \densS\,W|.
$
If $w \sim 1$, we say that the state $\densS$ is highly nonclassical. This implies that for all classical-like states $\densS$ the quantumness witness should be small. The inverse is not true: $w\ll 1$, does not necessarily imply that the state is classical-like. This construction may look rather abstract at this point. Later on we will consider an example which will help to clarify its meaning and merits.

Now we are in a position to formulate the
{\it extended eigenstate decoherence hypothesis}.
To this end we define a subset $\Delta$ of eigenstate labels which satisfies the condition
\be
\forall ~m,n\in \Delta:~~~ |E_m-E_n|\ll 2\pi/\tau
\ee
and introduce coefficients  $a_{mn}$, $m,n\in \Delta$ satisfying $a_{mn}=a_{nm}^*$ and $\sum_n |a_{nn}|^2=1$.
The hypothesis states that any state $\densS$ of the form
\be
\densS=\sum_{m,n \in \Delta} a_{mn}  \densS_{nm},
\ee
is classical-like, i.e. $w(\densS)\ll 1$.\\

The previous formulation of the EDH \cite{lychkovskiy2013dependence} comes as a particular case of the present one, with $a_{mn}=\delta_{m n}$.

The  extended EDH ensures that terms in eq. \eqref{densSt} with $|E_m-E_n|\ll 2\pi/\tau$ will not introduce quantumness in the state of the subsystem, despite the corresponding exponents do not average out on the time scale of decoherence,~$\tau$.

It can be illuminating to somewhat narrow the scope of the extended EDH and consider pairs of eigenstates and fixed coefficients $a_{mn}$. This way we can formulate, for example, the following version of the EDH:\\

\smallskip

\noindent
For all pairs of eigenstates such that $|E_m-E_n|\ll 2\pi/\tau$
\be\label{EDH vulgar}
w_{mn}=\frac12 \langle \Phi_m+\Phi_n | W | \Phi_m+\Phi_n \rangle \ll 1
\ee
It is this form of the extended ETH which we test below.

\medskip
\noindent{\it Example.} Here we consider a system $\cH$ of $N$ spins $1/2$ on a lattice with Heisenberg interaction,
\begin{equation}
\label{H}
H=\frac12\sum_{\langle i,j \rangle} \ssigma_i \ssigma_j,
\end{equation}
where $\ssigma_i$ is the sigma-matrix of the $i$'th spin, and summation is performed over pairs $\langle i,j \rangle$ determined by  some graph. For our numerical example we consider the graph with $N=10$ sites shown in Fig. \ref{fig}.

A subsystem $\cS$ contains first $M=5$ sites, while the rest $N-M=5$ sites belong to the environment $\cB$. We choose quantumness witness operator to be
\be\label{W}
W=\prod_{j=1}^M \sigma^+_j+ \prod_{j=1}^M \sigma^-_j,
\ee
where $\sigma^\pm_j=(\sigma_j^x\pm\ii \sigma_j^y)/2$. Such quantumness witness detects superpositions of total magnetization of the form
\be
\psiS=\frac1{\sqrt2}\left(|\uparrow\uparrow\uparrow\uparrow\uparrow\rangle+ |\downarrow\downarrow\downarrow\downarrow\downarrow \rangle\right).
\ee
In fact, this highly nonclassical state is an eigenstate of $W$, and
 $w\big(|\psiS\rangle\langle\psiS|\big)=\langle\psiS|W|\psiS\rangle =1.$

It is clear from the  numerical results shown in Fig. 1 that $w$ calculated according to eq. \eqref{EDH vulgar} is suppressed for small $w_{mn}=|E_m-E_n|$, in agreement with the extended EDH~\eqref{EDH vulgar}.

For comparison we show the matrix element for another operator,
\be\label{C}
C=\prod_{j=1}^M \sigma^z_j.
\ee
This operator has classical-like eigenstates and thus can not serve as a quantumness witness. One can see that in this case nothing particular happens around $|E_m-E_n|\simeq0$.

As a final remark, we note that decoherence is much less sensitive to integrability than thermalization. The reason is that the system is not able to explore whether it is integrable or not on the decoherence time scale. This implies that the extended eigenstate decoherence hypothesis should be applicable to  integrable systems without major modifications. We have tested that this is indeed the case in an integrable Heisenberg model.

\begin{acknowledgments}
{\it Acknowledgements.} The support from the Russian Science Foundation under the grant N$^{\rm o}$ 17-11-01388 is acknowledged.
\end{acknowledgments}

\bibliography{C:/D/Work/QM/Bibs/macrosuperpositions,C:/D/Work/QM/Bibs/decoherence,C:/D/Work/QM/Bibs/QIP,C:/D/Work/QM/Bibs/spin_chains,C:/D/Work/QM/Bibs/thermalization}

\end{document}